\documentclass[]{aa}
\usepackage{url}
\usepackage{natbib}               
\usepackage{graphicx}             
\usepackage{amssymb}              

\newcommand{\BE}{\begin{equation}}
\newcommand{\EE}{\end{equation}}
\newcommand{\BA}{\begin{align}}
\newcommand{\EA}{\end{align}}
 \newcommand{\fig}[1]{Figure~\ref{fig_#1}}
 \newcommand{\figS}[1]{Figures~\ref{fig_#1}}
 
 \newcommand{\figs}[2]{Figures~\ref{fig_#1} and \ref{fig_#2}}
 
 \newcommand{\sect}[1]{Section~\ref{sect_#1}}
 \newcommand{\sects}[2]{Sections~\ref{sect_#1} and~\ref{sect_#2}}
 \newcommand{\eq}[1]{Equation~(\ref{eq_#1})}
 \newcommand{\eqs}[2]{Equations~(\ref{eq_#1}) and (\ref{eq_#2})}
 \newcommand{\eqss}[2]{Equations~(\ref{eq_#1}) - (\ref{eq_#2})}
 
\newcommand{\eg}{e.g.,}

\newcommand{\ie}{i.e.,}
\newcommand{\insitu}{in situ}

\newcommand{\f}[2]{{\ensuremath{\mathchoice%
        {\dfrac{#1}{#2}}
        {\dfrac{#1}{#2}}
        {\frac{#1}{#2}}
        {\frac{#1}{#2}}
        }}}
\newcommand{\Int}[2]{\ensuremath{\mathchoice%
        {\displaystyle\int_{#1}^{#2}}
        {\displaystyle\int_{#1}^{#2}}
        {\int_{#1}^{#2}}
        {\int_{#1}^{#2}}
        }}

\newcommand{\rmd}{{\rm d }}

\newcommand{\kms}{km\,s$^{-1}$}

\newcommand{\der}[2]{\f{\rmd \, #1}{\rmd \, #2}}

\newcommand{\Bc}{B_{\rm c}}

\newcommand{\cB}[2]{c_{\rm #1,#2}}

\newcommand{\Do}{D_{0}}
\newcommand{\dt}{\Delta t}
\newcommand{\dVlin}{\Delta V_{\rm lin}}
\newcommand{\dVquad}{\Delta V_{\rm quad}}

\newcommand{\ffit}{f_{\rm fit}}
\newcommand{\fmod}{f_{\rm mod}}

\newcommand{\sdRes}{sd_{\rm residual}}

\newcommand{\tend}{t_{\rm end}}
\newcommand{\ti}{t_{\rm i}}
\newcommand{\tstart}{t_{\rm start}}
\newcommand{\tB}{t_{\rm B}}
\newcommand{\tref}{t_{\rm ref.}}
\newcommand{\tc}{t_{\rm c}}
\newcommand{\tj}{t_{\rm j}}
\newcommand{\Vcent}{V_{\rm cent.}}
\newcommand{\Vint}{V_{\rm j,\,int.}}
\newcommand{\Vj}{V_{\rm j}}
\newcommand{\Vfit}{V_{\rm fit}}
\newcommand{\Vobs}{V_{\rm obs}}
\newcommand{\Vc}{V_{\rm c}}
\newcommand{\Vmod}{V_{\rm mod}}
\newcommand{\xc}{x_{\rm c}}
\newcommand{\xcent}{x_{\rm cent.}}
\newcommand{\xint}{x_{\rm j,\,int.}}
\newcommand{\xj}{x_{\rm j}}
\newcommand{\xend}{x_{\rm end}}
\newcommand{\xstart}{x_{\rm start}}

\begin{document}

\title{Contribution of the aging effect to the observed asymmetry of interplanetary magnetic clouds}

\titlerunning{Aging effect contribution to the MC asymmetry}
\authorrunning{D\'emoulin et al.}

\author{P. D\'emoulin\inst{1,2}, S. Dasso\inst{3,4}, V. Lanabere\inst{4}, M. Janvier\inst{5}
\and C. No\^{u}s\inst{2} }
   \offprints{P. D\'emoulin}
\institute{
$^{1}$ LESIA, Observatoire de Paris, Universit\'e PSL, CNRS, Sorbonne Universit\'e, Univ. Paris Diderot, Sorbonne Paris Cit\'e, 5 place Jules Janssen, 92195 Meudon, France, \email{Pascal.Demoulin@obspm.fr}\\
$^{2}$ Laboratoire Cogitamus, 1 3/4 rue Descartes, 75005 Paris, France \\
$^{3}$ CONICET, Universidad de Buenos Aires, Instituto de Astronom\'\i a y F\'\i sica del Espacio, CC. 67, Suc. 28, 1428 Buenos Aires, Argentina, \email{sdasso@iafe.uba.ar} \\
$^{4}$ Universidad de Buenos Aires, Facultad de Ciencias Exactas y Naturales, 
Departamento de Ciencias de la Atm\'osfera y los Oc\'eanos and Departamento 
de F\'\i sica, 1428 Buenos Aires, Argentina, \email{dasso@df.uba.ar}\\
$^{5}$ Institut d'Astrophysique Spatiale, UMR8617, Univ. Paris-Sud-CNRS, Universit\'e Paris-Saclay, B\^atiment 121, 91405 Orsay Cedex, France \email{miho.janvier@u-psud.fr}\\
}

   \abstract  
   {Large magnetic structures are launched away from the Sun during solar eruptions.   They are observed as (interplanetary) coronal mass ejections (ICMEs or CMEs) with coronal and heliospheric imagers. A fraction of them are observed \insitu\ as magnetic clouds (MCs). Fitting these structures properly with a model requires a better understanding of their evolution.}
   {In situ measurements are done locally when the spacecraft trajectory crosses the magnetic configuration.  These observations are taken for different elements of plasma and at different times, and are therefore biased by the expansion of the magnetic configuration.   This aging effect leads to stronger magnetic fields measured at the front than at the rear of MCs, an asymmetry often present in MC data.  However, can the observed asymmetry be explained quantitatively only from the expansion? 
   }
   {Based on self-similar expansion, we derive a method to estimate the expansion rate from observed plasma velocity.  We next correct for the aging effect both the observed magnetic field and the spatial coordinate along the spacecraft trajectory.  This provides corrected data as if the MC internal structure was observed at the same time.
   }
   {We apply the method to 90 best observed MCs near Earth (1995-2012). The aging effect is the main source of the observed magnetic asymmetry only for 28\% of MCs. After correcting the aging effect, the asymmetry is almost symmetrically distributed between MCs with a stronger magnetic field at the front and those at the rear of MCs. 
   }
   {The proposed method can efficiently remove the aging bias within \insitu\ data
of MCs, and more generally of ICMEs.  This allows one to analyse the data with a spatial coordinate, such as in models or remote sensing observations. 
   }

    \keywords{Physical data and processes: magnetic fields, Sun: coronal mass ejections (CMEs), Sun: heliosphere 
    }

   \maketitle

\section{Introduction} 
\label{sect_Introduction}
Magnetic emergence and photospheric motions stress the coronal magnetic field, which at some critical point can become unstable.  In the frequent cases where the stable overlying magnetic field is not strong enough, the instability develops in to the ejection of plasma and magnetic field away from the Sun.  These events are called coronal mass ejections (CMEs) and are routinely observed by coronal imagers and coronagraphs \citep[\eg ][]{Sheeley85,Schwenn06,Howard11,Chen17}. 
  When an ejection is crossed by a spacecraft in the interplanetary space, local plasma and magnetic field measurements can be made. A variety of possible criteria has been defined to identify the ejecta signature in the data, and since they are different from the remote-sensing observations of CMEs, the crossed ejections are called interplanetary CMEs \citep[ICMEs, \eg ][]{Cane03,Wimmer-Schweingruber06,Zurbuchen06,Demoulin10b,Kilpua17b}.  
  When available, the heliospheric imagers can provide a link between a CME observed close to the Sun and an ICMEs observed \insitu , showing that ICMEs are the continuation of CMEs away from the Sun \citep[\eg ][]{Harrison09,Rouillard11b,Mostl14}.
An ICME is typically formed at their front by compressed plasma and magnetic field (the sheath) followed by a magnetic ejecta \citep[\eg ][]{Cane97,Winslow15,Janvier19} which is thought to be the continuation of the coronal eruption.  When the ejecta has a smooth, coherent and large rotation of the magnetic field, and a low proton temperature (compared to the typical solar wind at the same speed), it is classified as a magnetic cloud (MC) \citep[\eg ][]{Burlaga81,Gosling90}.
Such magnetic structures are typically modelled with twisted magnetic flux tubes, or flux ropes \citep[FRs, \eg ][]{Lepping90,Lynch03,Dasso09b,Lanabere20}.

 Coronagraph observations allow us to follow CMEs, showing that they typically expand nearly proportionally with solar distance, $D$, away from the low corona (\ie\ the size $S$ increase as $S \propto D^{\zeta}$ with $\zeta \approx 1$).  In contrast, \insitu\ data are typically available only at one solar distance per event since coalignement, within few degrees, of two spacecraft observing the same event at different solar distances are rare \citep{Nakwacki11,Vrsnak19,Good19}.  Then, results are typically obtained by analysing large sets of events supposing that the statistic is large enough at each distance to erase the individual properties.  Such statistical analyses show that the MC radial size increases as a power law of the solar distance \citep{Kumar96,Bothmer98,Leitner07,Gulisano10}.   
These results were extended to ICMEs \citep{Liu05,Wang05b}.  The exponent $\zeta$ is around unity, so the ICME size is nearly proportional to the solar distance, $D$, with variations depending both on the sample and the distance range analysed as summarised in Table 1 of \citet{Gulisano12}.

The \insitu\ velocity temporal profiles provide a more direct and systematic way to access the expansion properties within individual events.  Indeed, the proton velocity profile typically decreases almost linearly during the spacecraft passage through the MC \citep{Lepping03b, Lepping08, Jian08c, Gulisano10, Gulisano12, Rodriguez16, Masias-Meza16}. This kind of velocity profile is expected for a self similar expansion \citep{Shimazu02,Demoulin08}. Recently, such self similar expansion was found compatible with the magnetic field profiles obtained for 18 ICMEs observed twice with two spacecraft nearly radially aligned from the Sun \citep{Good19,Vrsnak19}.
 The physical origin of such an evolution is due to the steep decrease of the solar wind total pressure that follows a power law of the solar distance.   An approximative pressure equilibrium of the ICME with its surrounding solar wind induces an expansion factor that is also governed by a power law \citep{Demoulin09}.   
 
A general theoretical framework of the expansion was developed with a hierarchical order from the most general case (ICMEs, anisotropic expansion) to more specific ones (\eg\ FRs), with the specification of the context and hypothesis at each step \citep{Demoulin08}. The proton velocity profile measured along the spacecraft trajectory allows us to estimate the expansion rate $\zeta$.  These local results are globally agree with the statistical results of ICME and MC size evolution versus solar distance \citep{Bothmer98, Gulisano10, Gulisano12}.  

The expansion affects the \insitu\ measurements since during the spacecraft crossing the magnetic structure has evolved.  This is known as the aging effect. For a configuration in expansion this implies a bias with a stronger magnetic field measured at the front, when the configuration was smaller, than later on when measured at the rear.  FR models are typically developed with static configurations, then a fit of such models to MC data introduces a bias in the derived parameters.  To overcome this, a self similar expansion is typically supposed.  This introduces extra free parameters which are found by including in the fitting procedure the observed proton velocity.  The earlier attempts supposed only a 2D expansion orthogonal to the FR axis \citep{Farrugia93,Osherovich93,Nakwacki08}.  However, this approach is unphysical since the magnetic configuration becomes force-imbalanced during the evolution.  A more consistent solution is to suppose an isotropic 3D expansion.  The parameters of the magnetic and velocity models are found by minimising the deviation from the model to both velocity and magnetic data combined within a single function  \citep{Shimazu02,Dasso07,Vandas06,Marubashi07,Lynnyk09}. 

The improvements of the fit of the magnetic field components by a model in expansion
for specific MCs \citep[\eg ][]{Vandas15b,Marubashi17,Vandas17b} could be an indication that the aging effect is the main origin of the frequently observed stronger magnetic field found in the front of MCs compared to the values at their rear. However, this is likely to be not so general since this asymmetry is only present in a fraction of MCs, typically faster ones \citep{Masias-Meza16}.  Furthermore, MCs having a reverse velocity profile, so in compression, are also observed \citep[\eg ][]{Gulisano10}.   

The main aim of this paper is to further develop the measurement of MC expansion from \insitu\ data, then to remove its effects on the magnetic field data. 
In \sect{Exp} we present the equations describing the self-similar expansion of ICMEs
which are relating the expansion factor to the observed velocity.  We derive a methodology to apply these equations to \insitu\ data.  In \sect{Exp_Profiles} we derive the expansion profiles of well observed MCs, then in \sect{Exp_Asy} we investigate if the aging effect can explain the observed magnetic asymmetry.  Next, in \sect{Rem} we present a procedure to correct \insitu\ magnetic data for the aging effects.  Finally, in \sect{Conclusion} we summarise our results and conclude.

\section{Derivation of expansion rate from velocity profile}
\label{sect_Exp}

\subsection{Spatial coordinate and aging effect}
\label{sect_Exp_x(t)}

The \insitu\ measurements provide plasma and magnetic data as a function of time, $t$, as the moving magnetised plasma is crossed by the observing spacecraft. A blob of plasma moving at the velocity $\Vobs(t)$ and observed around time $t$ during $\rmd t$ has a spatial extension     
  \BE \label{eq_dx(t)}
  \rmd x(t) = \Vobs(t) \,\rmd t \,
  \EE 
Then, a spatial coordinate along the spacecraft trajectory is derived by integrating the velocity component $\Vobs(t)$ observed along the trajectory as
  \BE \label{eq_x(t)}
  x(t) = \Int{\tref}{t} \Vobs(\ti) \,\rmd \ti  \,,
  \EE 
where $\tref$ is a selected reference time.  Compared to the original data, provided as a function of time, rewriting the data as a function of the spatial coordinate $x$ corrects the effect of the velocity (\eg\ a faster blob of plasma appears shorter in the original data).  

However, $x(t)$ is not a true spatial coordinate as the crossed structure typically changes its size during the spacecraft crossing.
For example, since the MCs shown in \figS{Vfit}a-d are in expansion, the MCs grow in size with time.  This global expansion affects each plasma blob whose size $\rmd x$ grows with time.  
Then, $\rmd x$ is larger when the plasma blob is observed later in the MC.  For an MC in contraction, the reverse, \ie\ a smaller size, is deduced from \eq{dx(t)} when observations are taken closer to the MC rear. 
  This observational bias is called the aging effect. Indeed, \insitu\ observations mix time and space since the intrinsic temporal evolution of a plasma blob cannot be observed.  Rather a plasma blob is observed only once, and a different plasma blob is observed at each time.
We conclude that $\rmd x(t)$ in \eq{dx(t)} first needs to be corrected for the aging effect, then a time integration could provide a true spatial coordinate at a given time across the observed structure (such as in remote sensing observations).  In the following, we derive a method to compute the expansion or compression temporal profile from observed velocities.  This is later used to correct the aging effect in the observed data (\sect{Rem}).

\subsection{Global and internal motions}
\label{sect_Exp_Gen}

    We present this study in the framework of the 3D expansion theory of ICMEs described by \citet{Demoulin08}, simplifying it to the minimum amount needed to analyse the data within ICMEs crossed by a single spacecraft.  
Such data are the most numerous presently available. The equations derived by \citet{Demoulin08} for self-similar expansion with possible different rates in three orthogonal directions are only partially constrained by the data of a single spacecraft.
Still, since the plasma data constrain well the expansion rate along the spacecraft trajectory, we develop this approach below in order to extract the most possible informations from the data.  It applies to ICME intervals observed in global expansion (or contraction), in particular to MCs.

We first suppose that the motion of a plasma blob could be described by the sum of a global motion and an internal motion.   The global motion is affected by the external forces such as the drag force.   The internal motion is driven by internal forces such as the imbalance of total pressure both in the MC volume and at its boundary with the surrounding medium.   Setting an axis coordinate $x$ along the spacecraft trajectory, with its origin fixed at the Sun, the location of a plasma blob, labeled j, writes 
  \BE \label{eq_xj(t)}
  \xj(t) = \xcent(t) + \xint(t) \,,
  \EE
where $\xcent(t)$ is the position of the centre of the studied event and $\xint(t)$ is the relative position of blob j with respect to the centre. $\xcent(t)$ describes the center of mass motion and $\xint(t)$ the internal evolution.
The time derivative of \eq{xj(t)} provides the plasma blob velocity
  \BE \label{eq_Vj(t)}
  \Vj(t) = \Vcent(t) + \Vint(t) \,.
  \EE

 Ideally, the centre is the centre of mass, so that its motion is governed by the resultant of all external forces applied to the MC.   In practice, with \insitu\ data, this mass centre cannot be determined since only a 1D cut is available. The approximation available from \insitu\ data only allows us to set the MC centre at the time $\tc=(\tstart +\tend)/2$ where $\tstart$ and $\tend$ are defined at the MC boundaries. 
 
We next suppose that the internal evolution is a self-similar expansion (or contraction) in the $x$-direction.  This implies that the configuration at a time $t$ is a scaled version of the configuration present at another time $\tc$, which writes as 
  \BE \label{eq_xint(t)}
  \xint(t) = \xint(\tc) \, f(t) \,.
  \EE
where we set the expansion factor $f(\tc)=1$ at the reference time $\tc$.  
In other words, $\xint(\tc)$ is a Lagrangian coordinate of the followed plasma blob j.
Thus, the internal velocity can be expressed as $\Vint(t) = \rmd \xint(t) / \rmd t = \xint(\tc) \, \rmd f(t)/ \rmd t$.  Then, \eq{Vj(t)} reads  
  \BE \label{eq_Vj2(t)}
  \Vj(t) = \Vcent(t) + \xint(\tc) \,\, \der{f(t)}{t} \,.
  \EE

During an MC crossing, the observing spacecraft samples different plasma blobs j at times noted $\tj$.  The crossing time is typically short, about one day, compared with the time scale of the spacecraft trajectory evolution (about one year for a spacecraft located at 1 au from the Sun).  Then, the spacecraft is approximately at a fixed distance from the Sun, called $\Do$ below \citep[see more justifications in Section 2.1 of][]{Demoulin08}.
Then, including \eq{xint(t)} in \eq{xj(t)} with $t=\tj$, \eq{xj(t)} is rewritten as
  \BE \label{eq_xj(tj)}
  \xj(\tj) = \xcent(\tj) + \xint(\tc) \, f(\tj) = \Do \,. 
  \EE
This provides an expression for the unknown position $\xint(\tc)$ as
  \BE \label{eq_xint(tc)}
  \xint(\tc) = (\Do-\xcent(\tj) ) ~/f(\tj) \,,
  \EE  
allowing us to eliminate $\xint(\tc)$ in \eq{Vj2(t)} written at $t=\tj$. It rewrites     
  \BE \label{eq_Vj(tj)}
  \Vj(\tj) = \Vcent(\tj) +  (\Do-\xcent(\tj) )  \, \left(\der{\ln f(t)}{t}\right)_{t=\tj} \,.
  \EE
Above we mark explicitly the plasma blob with the index j for the derivation of the equations making a difference between following a given plasma blob j with time $t$, \eqss{xj(t)}{Vj2(t)},
from observing different plasma blobs j at the spacecraft position and at different times $\tj$, \eqss{xj(tj)}{Vj(tj)}.
However, this derivation being achieved, we can simplify the final equation 
by changing $\tj$ to $t$ in \eq{Vj(tj)}, which expresses the continuous observations at the spacecraft with time $t$. This implies that below we keep implicit the reference to observations of different plasma blobs with time and only refer to the observed velocity with the index "obs". 
In conclusion, the observed velocity in MCs, \eq{Vj(tj)}, is generically modelled as
  \BE \label{eq_Vobs(t)}
  \Vobs(t) = \Vcent(t) +  (\Do-\xcent(t) )  \, \der{\ln f(t)}{t} \,.
  \EE

\subsection{Constraints on the global motion}
\label{sect_Exp_Cons}

The observed proton velocity, $\Vobs(t)$, has two contributions: first the global motion (described by $\Vcent(t)$) and the expansion (described by $f(t)$).   These two contributions cannot be separated within the data without extra information. 

 The often observed decreasing profile of $\Vobs(t)$ with time in ICMEs could a priori be a consequence of a global deceleration of the ICME when it encounters the spacecraft.   However, the magnitude needed for this deceleration would be one to two orders of magnitude larger than the estimated deceleration obtained with three independent methods, as follows.

  A first method to estimate the acceleration far from the Sun consists in using observations in quadrature with a coronagraph imaging the core of the CME while the same event is observed \insitu\ with another spacecraft. These two observations provides an estimation of a mean acceleration from the Sun to the spacecraft, which is an upper bound of the acceleration at the crossing spacecraft since acceleration is stronger close to the Sun \citep[\eg ][]{Rust05}.  
  A second method consists in deriving statistically the dependence of the ICME velocity with solar distance from \insitu\ data taken at different solar distances \citep[\eg ][]{Liu05}, then to deduce a typical acceleration.
  The third method uses observations of spacecraft nearly radially aligned from the Sun and observing the same ICME at different solar distances \citep[\eg ][]{Cane97, Good19, Salman20}. 
The velocity measurements at both spacecraft, the distance to the Sun and the timing at both spacecraft, allow us the derivation of two independent estimations of the mean acceleration. 
These three independent methods show that the acceleration of the ICME center is generally too weak to explain the profiles of \insitu\ velocities  \citep[see][for a quantitative analysis which could be updated with the confirmation obtained with the above more recent results]{Demoulin08}.  

    The above result was confirmed by studies with the imager data of the SoHO and STEREO spacecraft \citep{Liu16,Wood17,Zhao19}.  Indeed, most of the CME deceleration occurs close to the Sun, and the faster events have a stronger deceleration confined closer to the Sun. 
The imager data are typically consistent with a constant CME velocity for distances from the Sun above 0.3 au, and for all cases above 0.6 au.   Then, the results of the imager data imply that the variations of $\Vcent(t)$ cannot explain observed \insitu\ velocity variations across ICMEs, and in particular within MCs, which are typically between 50 and 100 \kms .   

\subsection{Expansion rate derived from \insitu\ observations}
\label{sect_Exp_insitu}

In the line of the results of the \insitu\ and imager data on ICMEs and CMEs on their way to 1 au, as summarised above, we suppose a constant MC velocity, $\Vc$, of its centre during the spacecraft crossing, 
  \BE \label{eq_xc(t)}
  \xcent(t) = \Vc (t-\tc) + \Do \,.
  \EE 
Then, \eq{Vobs(t)} is rewritten as  
  \BE \label{eq_Vobs(t,f)}
  \Vobs(t) = \Vc - \Vc (t-\tc)  \, \der{\ln f(t)}{t} \,. 
  \EE  
This provides a direct link between the expansion factor $f(t)$ and the observed velocity $\Vobs(t)$, which could be rewritten as
  \BE \label{eq_f(t,Vobs)}
  \der{\ln f(t)}{t} = -\frac{\Vobs(t)-\Vc}{\Vc\ (t-\tc)} \,. 
  \EE 
Then, \eq{f(t,Vobs)} shows that the temporal derivative of the logarithm of $f(t)$ can be calculated with the finite difference of the observed velocity profile computed at $t$ and $\tc$ and normalised with $\Vc$.  Supposing $\Vobs(t)>\Vc$ for $t<\tc$, and the reverse for $t>\tc$ (case in expansion), then $\rmd f(t)/ \rmd t >0$ which implies that the lowest and largest $f(t)$ values are expected at the front and rear MC boundaries, respectively.  The same conclusion applies, with an exchange of extrema between boundaries for $\Vobs(t)<\Vc$ for $t<\tc$, and the reverse for $t>\tc$ (case in compression).
  
The previous formalism allows us to derive the expansion factor evolution with time, $f(t)$, directly  from the observed velocity $\Vobs(t)$ by integrating \eq{f(t,Vobs)} as
  \BE \label{eq_f(t)int}
  f(t) = \exp \Bigg( \Int{\tc}{t} \f{1-\Vobs(\ti)/\Vc}{\ti-\tc} \, \rmd \ti \Bigg)  \,,
  \EE  
with $f(\tc)$ set to unity (so that $\tc$ is the reference time of the magnetic configuration). 
The integrant is undetermined for $\ti=\tc$ since both the denominator and the numerator vanish.  However, if $\Vobs(\ti)$ is derivable, a first order Taylor expansion of $\Vobs(\ti)$ around $\ti=\tc$ removes this indetermination.  More precisely, this difficulty disappears with $\Vobs(\ti)$ written as $\Vc + (\ti-\tc)\, (\rmd V/\rmd \ti)|_{\ti=\tc} + (\ti-\tc)^2 \, U(\ti)$, where $U(\ti)$ contains the rest of the expansion and is finite.    

The needed smoothness of $\Vobs(\ti)$ in integrating \eq{f(t)int} could be achieved with a local polynomial interpolation around $\ti=\tc$ of $\Vobs(\ti)$ data (\eg\ with a spline interpolation).  This approach has the advantage of incorporating directly the data in the computation of $f(t)$.   However, several phenomena, such as waves and local flows, have contributions in the observed $\Vobs(t)$ profile.  Such phenomena cannot be modelled with a self-similar expansion hypothesis.  Then, we choose to filter-out all the velocity contributions at scales smaller than the MC size by first performing a polynomial fit of the observed $\Vobs(t)$.  Results on MCs show that it is not worth to go beyond a polynomial of second order (see \sect{V_Profiles}),
  \BE \label{eq_Vfit}
  \Vfit (t) = a + b\, (t-\tc) + c\, (t-\tc)^2 \,,
  \EE  
where $a,\, b,\, c$ are the fitted coefficients to the data.  
The coefficient $a$ is the estimated velocity at the MC centre ($t=\tc$), and
the coefficients $b$ and $c$ describe the expansion.   

After inclusion of \eq{Vfit} in \eq{f(t)int}, the integration provides 
  \BE \label{eq_fFit}
  \ffit (t) = e^{-\f{b}{a}(t-\tc)} \, e^{-\f{c}{2a}(t-\tc)^2} \,,
  \EE  
which is a well behaving function of $t-\tc$.

\subsection{Expansion rate derived from an expansion model}
\label{sect_Exp_model}

In parallel to the previous approach based on fitting the velocity data obtained at a fixed solar distance, $D$, we explore below another approach based on studies analysing MC sizes observed at various solar distances.  
These statistical studies typically found a power law dependence of the MC size with solar distance (see \sect{Introduction}). 
It was shown theoretically that such dependence is expected from the observed power law decrease of the total plasma pressure of the solar wind with distance \citep{Demoulin09}.  
These results imply that $f(t)$ is typically expected to be a power law of solar distance $\xcent(t)$,
  \BE \label{eq_fPLaw}
  \fmod (t) = \bigg(\f{\xcent(t)}{\Do} \bigg)^{\zeta} \,,
  \EE  
with $\Do=\xcent(\tc)$ being included to have the same normalisation as above ($\fmod (\tc) =1$).  
The average variation of the total pressure in the solar wind with distance determines a typical $\zeta$ value. 
However, here we want to analyse individual MCs where the total pressure in the surrounding solar wind and its variation with distance are not observed.   Then, we let $\zeta$ to be a free coefficient which is determined from the MC \insitu\ data. 

Like in \sect{Exp_insitu}, we suppose a constant velocity, $\Vc$, for the MC centre during the spacecraft crossing.  Including \eq{xc(t)} in \eq{fPLaw}, $\fmod (t)$ is rewritten as
  \BE \label{eq_fmod}
  \fmod (t) = \left(1 +  \f{\Vc\ (t-\tc)}{\Do} \right)^\zeta \,.
  \EE  
Including this expansion rate in \eq{Vobs(t)}  provides a model for $\Vobs(t)$ as
  \BE \label{eq_Vmod}
  \Vmod (t) = \Vc  - \f{\zeta \,\Vc^2 \,(t-\tc)/\Do}{1+\Vc\, (t-\tc)/\Do} \,.
  \EE  

When applied to the data, $\Do$ is the distance of the spacecraft to the Sun, and $\tc$ is the time when the center of the MC is observed. $\Vc$ and $\zeta$ are free parameters which can be determined by a least square fit of \eq{Vmod} to the velocity data.  

We next estimate the magnitude of the terms in \eq{Vmod}. For that we suppose that the crossed MC is formed of a FR having locally a cylinder shape of radius $R$ and with a FR axis inclined by an angle $\gamma$ on the spacecraft trajectory ($x$ axis).
With $\tB$ being the crossing time of one of the FR boundaries, $\Vc\, |\tB-\tc|$ is lower than $R /\sin(\gamma)$, the equality being obtained in the case of the spacecraft crossing the FR axis.  At the Earth orbit distance, typically $R/\Do \approx 0.1$ for MCs, with $R/\Do$ reaching rarely $\approx 0.2$ \citep[\eg\ ][]{Lepping90,Lepping15b}.        
Then, except for MCs crossed nearly along their axis or exceptionally large events, this implies that the denominator in \eq{Vmod} is nearly unity, which implies
  \BE \label{eq_VmodLinear}
  \Vmod (t) \approx \Vc  - \zeta \,\Vc^2 \,(t-\tc)/\Do \,.
  \EE  
Such a model describes the nearly linear velocity profile present in unperturbed MCs with the fit of \eq{VmodLinear} to data implying $\zeta \approx 1$ for the inner heliosphere \citep[HELIOS spacecraft,][]{Gulisano10}, to 1~au \citep[Wind and ACE spacecraft,][]{Demoulin08}, and even to the outer heliosphere up to 5 au  \citep[ULYSSES spacecraft,][]{Gulisano12}. 

Using the same linear approximation in \eq{fmod}, implies $\fmod (t) \approx 1+\zeta \,\Vc \,(t-\tc)/\Do$. Comparing to \eq{VmodLinear}, this implies a direct link between a linear approximation of the velocity and the expansion factor as $\fmod (t) \approx \Vc /\Vmod (t)$.   This provides a simple estimation of the expansion factor in MCs when the observed velocity is approximately linear (\sect{V_Profiles}).  In particular, at the MC boundary the expansion factor is about $f_{\pm} \approx 1 \pm \zeta \, R/\Do$ where $-$ and $+$ stands for the front and rear boundaries, respectively.  Taking $\zeta \approx 1$ and $R/\Do \approx 0.1$ provides an expansion factor between 0.9 and 1.1 in typical MCs.

\subsection{Comparison of the methods}
\label{sect_Exp_Comparison}

The two analyses in \sects{Exp_insitu}{Exp_model} have different hypotheses in order to apply the theory to MC observations.  They lead to different expressions for the velocity and the expansion factors.  In particular, the first method includes a quadratic fit of velocity observations and it expresses $f(t)$ with exponentials, \eq{fFit}, while the second method supposes that the MC size, so $f(t)$, is a power law of solar distance, \eq{fPLaw}.  This implies different velocity expressions, \eq{Vfit} and \eq{Vmod} respectively, with a different input of the data to determine the free parameters.
 
However, for practical applications to MCs, the two above methods are comparable, as follows.  The terms in $t-\tc$ have a low contribution compared to the leading terms due to the typically small radius of the FRs compared with the solar distance $\Do$, as introduced above before \eq{VmodLinear}.   We derive below a Taylor expansion to the second order in $t-\tc$ of the above equations in order to compare the two methods in the context of application to MCs.   

\eq{Vfit} is already a Taylor expansion to the second order of $\Vfit(t)$.  The same Taylor expansion applied to \eq{Vmod} is
  \BE \label{eq_VmodQuadra}
  \Vmod (t) \approx \Vc  - \zeta \,\Vc^2/\Do \,(t-\tc)
                               + \zeta \, \Vc^3/\Do^2 \,(t-\tc)^2 \,.
  \EE
Comparing this equation to \eq{Vfit}, implies 
  \BE \label{eq_c_mod}
  c=-a\,b /\Do\,,
  \EE
so that the model of \sect{Exp_model} can be considered as a particular case of the approach of \sect{Exp_insitu} when applied to MCs.  The Taylor expansion could also be applied to the expansion factors, \eqs{fFit}{fPLaw}.  With the same velocities, so using \eq{c_mod}, the equations are identical, as expected.

\section{Expansion profiles of MCs} 
\label{sect_Exp_Profiles}

    In this section, we apply the above theoretical description to observed MCs in order to derive the expansion factor $f(t)$ from \insitu\ plasma data.
 
\subsection{Analysed magnetic clouds}
\label{sect_Analyzed_MCs}

    We analyse the data obtained nearby Earth with the Wind spacecraft, and more precisely from the Magnetic Field Instrument (MFI) and the Solar Wind Experiment (SWE).  The data used have a temporal cadence of 60-sec for (MFI) and 92-sec for (SWE) and were downloaded from \url{https://cdaweb.sci.gsfc.nasa.gov/pub/data/wind/mfi/mfi_h0} and \url{https://cdaweb.sci.gsfc.nasa.gov/pub/data/wind/swe/swe_h1/} respectively.
The data are provided in the geocentric solar ecliptic (GSE) system of reference.

The 90 MCs included in Lepping's table with the best qualities (1 and 2) are analysed (\url{https://wind.gsfc.nasa.gov/mfi/mag_cloud_S1.html}).  This table summarises the results of the fitting of the magnetic data with the Lundquist model as described by \citet{Lepping90,Lepping10,Lepping15b}. These MCs observed by Wind between 1995--2012.

\subsection{Velocity profiles}
\label{sect_V_Profiles}

\begin{figure}[t!]        
\centering
\includegraphics[width=0.5\textwidth, clip=]{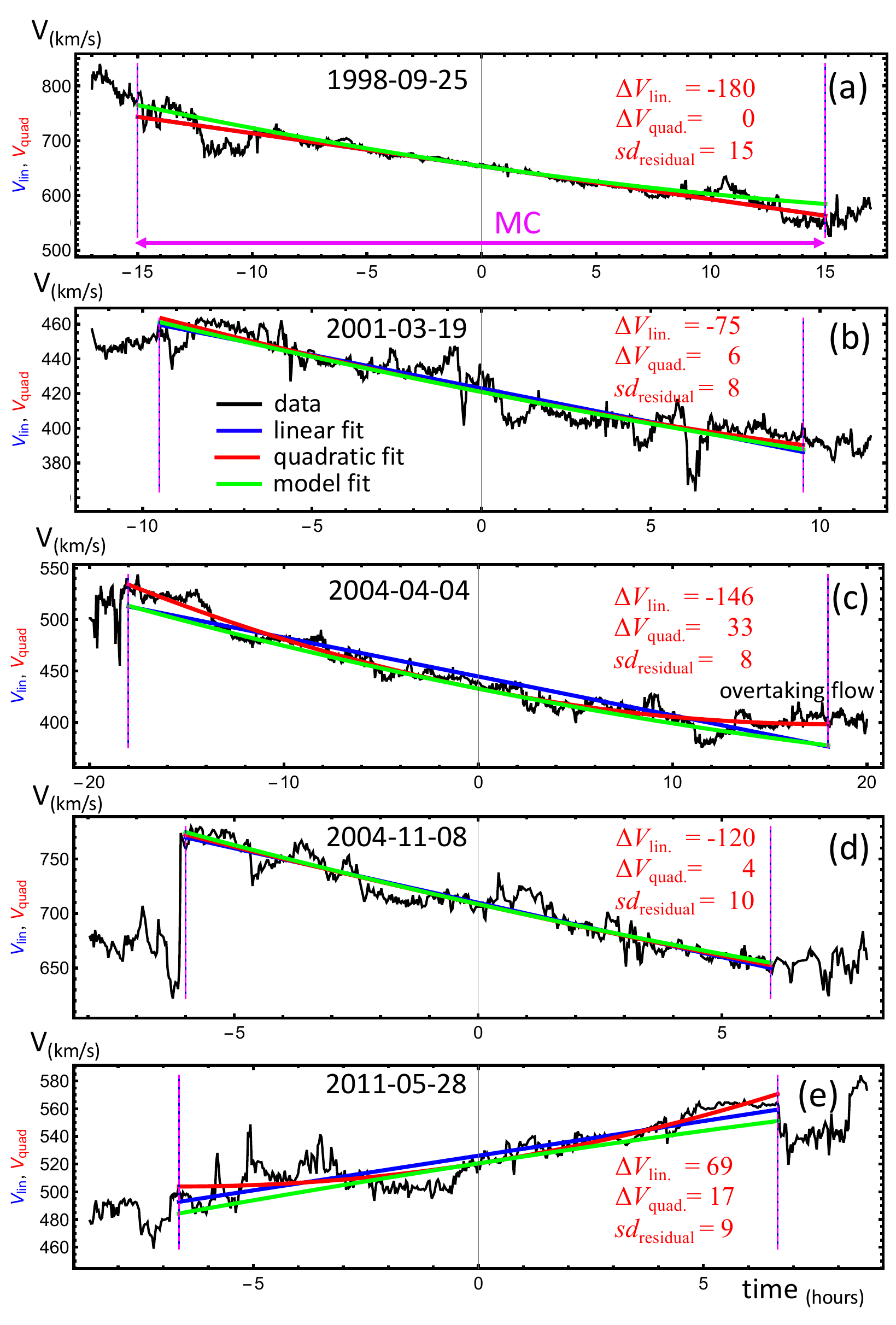}
\caption{ 
Examples of velocities measured \insitu\ (black) and fitted results (colours) for five MCs. The labels in the top of the plots indicate the day of the MC front. 
Linear (in blue) and quadratic (in red) least square fits, \eq{Vfit}, of the observed velocity are over-plotted.  The model fit (in green) is an approximation of
\eq{Vmod} (Taylor expansion to second order, so \eq{VmodQuadra}).
   The purple vertical lines define the boundaries of the MCs and the range of the velocity fits.  
Two hours of data are added before and after these boundaries to show the MC context.  
The time origin is set at MC centre, \ie\ $\tc=(\tstart +\tend)/2$.
The main characteristics of the quadratic fit are added in red font; all are in \kms 
(see \sect{V_Profiles}).
}
 \label{fig_Vfit}
\end{figure}

Five examples of MCs are shown in \fig{Vfit} with the velocity data (in black) and three fits (colours).  
The linear (in blue) and quadratic (in red) fits are derived from a least square fit of \eq{Vfit} to the data within the MC time interval (no expansion model is involved).  A power law model of the expansion with the solar distance, \eq{fmod}, is represented by the green line.  
More precisely, a Taylor expansion to second order of the derived velocity, \eq{VmodQuadra}, is used. 
The central time of the MC is set at $\tc =0$.

 The MC examples were selected to represent the variety of magnetic and velocity profiles while still being typical and well behaved MCs.  \figS{Vfit}a-d are examples of MCs in expansion, such as the large majority of MCs, and \figS{Vfit}e is an example of a MC in compression. Among these five examples,
two are large MCs (\figS{Vfit}a,c) with a long duration of $\dt \approx$ 27 and 37 h, respectively, and with a FR radius $\approx 0.2$ au from Lepping's results.  The three others, \figS{Vfit}b,d,e, have shorter duration of $\dt \approx$ 19 h, 12h and 13 h, respectively, and a smaller FR radius, 0.08, 0.09 and 0.06 au, respectively.  Two MCs are fast with $\Vc \approx 700$ \kms (\figS{Vfit}a,d), two are slow ones with $\Vc \approx 430$ \kms (\figS{Vfit}b,c) and the last one is intermediate with $\Vc \approx 520$ \kms (\figS{Vfit}e).   Next, apart from velocity fluctuations, three MCs have nearly linear velocity profiles (\figS{Vfit}a,b,d) and two have small deviation to linearity (\figS{Vfit}c,e).
   
The velocity profiles in \fig{Vfit} show mainly a linear variation with time across the MCs, which is expected from a global expansion or compression. 
Deviations to a linear profile are generally present with velocity fluctuations as well as when an overtaking flow is present, \ie\ when a fast solar wind stream impacts the MC rear, \eg\ like in \fig{Vfit}c.  In such a case, a moderate difference is present between the linear and quadratic fits. 
The deviation to a linear velocity profile is given by the third term in the righthand side of \eq{Vfit}.  This deviation vanishes at the MC center ($t= \tc$) and is maximal at both MC boundaries with values equal to $\dVquad = c \,(\dt/2)^{2}$,
while the linear variation of velocity across the MC is $\dVlin = b\, \dt$, where $\dt$ is the MC duration, while $b$ and $c$ are the coefficients of \eq{Vfit} fitted to the velocity data.  We also compute the standard deviation of residuals, $\sdRes$, between the velocity fits and the data.  The examples shown in \fig{Vfit} indicate that the linear change of velocity, $|\dVlin|$, is the dominant effect. 
  
  For 12 MCs, on a total of 90, a fast overtaking stream is present at the MC rear.  Then, the observed velocity jumps to a moderately higher value close to the MC rear.  The time interval with an enhanced velocity is short, below 6 hours, and on average only 2 hours.  In some MCs, this jump is likely the trace of a shock.  The correction of such effect would require applying the technique developed by \citet{Wang18} to remove the effect of the shock.  Here, we rather explore two options, either we fit the whole MC interval, either we remove the time interval after the shock for the velocity fit, so that the fit is closer to the observed velocity in most of the MCs.  A sudden jump in the observed velocity is also present for 5 MCs close to the front boundary, again within a short time interval, below 4 hours, and on average only 2 hours. 
Then, removing or not the above intervals has only a small effect on the results of a small fraction (19\%) 
of MCs, so that the statistical results below are weakly affected (changing the type of velocity fits is more important).

The above method is a different strategy than in our earlier papers where the aim was to compute the expansion factor $\zeta$ from the part of the velocity profile which was the closest from self-similar expansion so mostly linear with time \citep{Demoulin08,Gulisano10,Gulisano12}.
In our present work, our aim is rather to include the whole MC duration to compare the results with the three velocity fits, then to correct the magnetic data from expansion over the whole MC (or at least most of it).

\begin{figure*}[t!]        
\sidecaption
\centering
\includegraphics[width=0.6\textwidth, clip=]{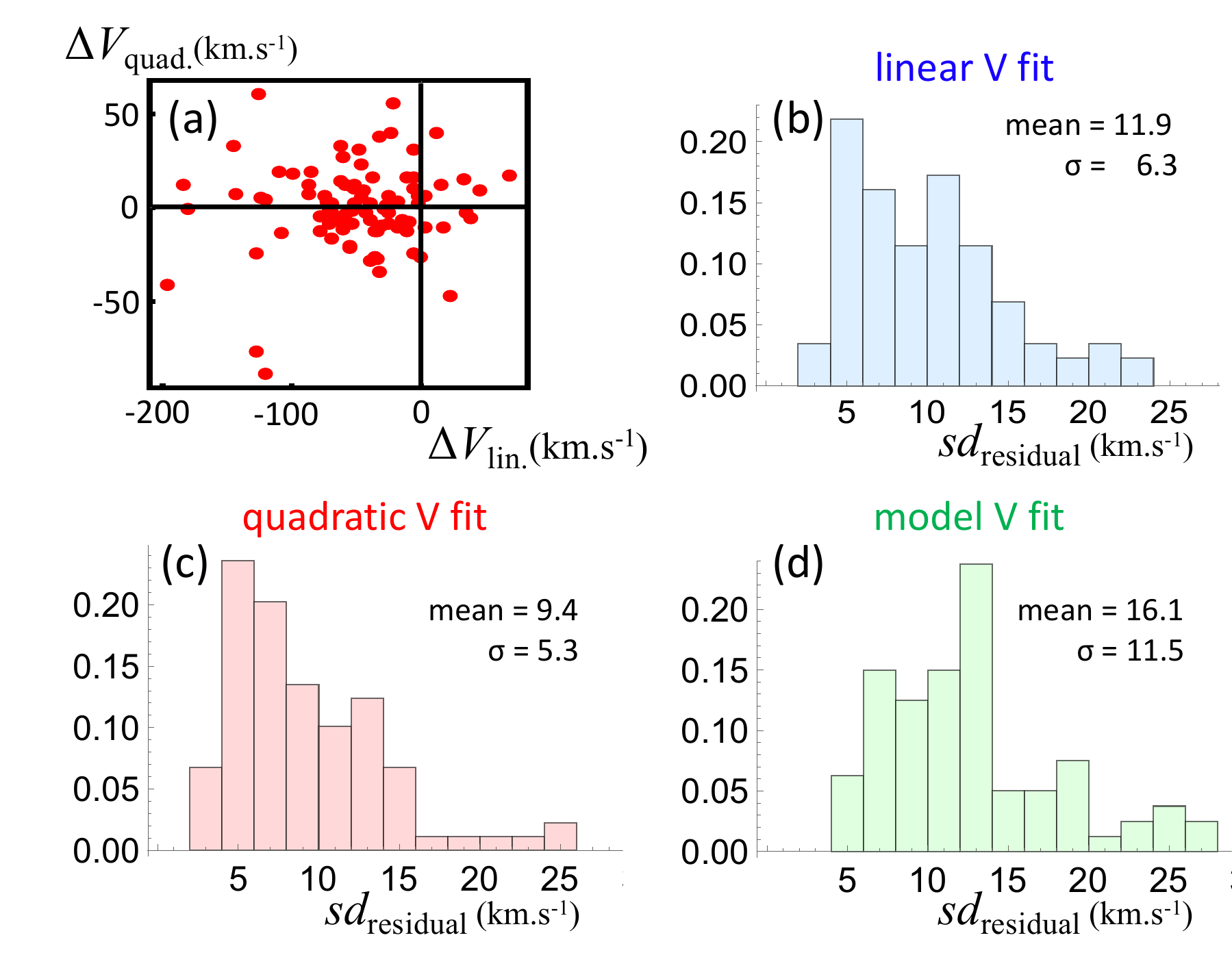}
\caption{Fit properties for the 90 MCs studied. 
(a) Maximum contribution across the MCs of the quadratic term, $\dVquad = c\, (\dt/2)^{2}$ as a function of the linear variation of velocity $\dVlin = b\, \dt$. $\dt$ is the MC duration, while $b$ and $c$ are the coefficients of \eq{Vfit} fitted to the velocity data. 
(b-d) Histograms of the standard deviation of residuals between the velocity fits and the data. The linear, quadratic, and model fits are in light blue, red, and green respectively.  
}
 \label{fig_dV}
\end{figure*}

   In \fig{dV}a the quadratic deviation $\dVquad$ to the linear fit is compared to the linear variation of velocity across the MC, $\dVlin = b \,\dt$.  The range of variation of $\dVlin$,  $\approx 300$ \kms, and its standard deviation, $\approx 50$ \kms,  are about twice the ones of $\dVquad$ ($\approx 150$ and $\approx 23$ \kms, respectively).  The dominance of negative  $\dVlin$, with a mean value of $\dVlin = -49 \pm 50$ \kms, implies that MCs are typically in expansion, while the quadratic term $\dVquad$ is nearly symmetrically distributed around the origin, with a mean value of $\dVquad = 0 \pm 23$ \kms .   There is no correlation between these two terms (the Pearson and Spearman correlation coefficients are 0.1 and 0.0 respectively).

Next, we investigate how far from the regression curve data points are by computing 
the standard deviation of the residuals, $\sdRes$ (\fig{dV}b-d).  The linear fit provides typically a rather fair fit of the velocity data as $\sdRes \approx 12 \pm 6$ \kms , with a maximum of 23 \kms . As expected with one more free parameter, the quadratic fit has lower residuals, $\sdRes \approx 9 \pm 5$ \kms . 
   Next, the model fit is derived from the quadratic fit with the coefficient $c$ imposed by \eq{c_mod}. This increases significantly the deviation to the data ($\sdRes \approx 16 \pm 12$ \kms ).  This implies that the global expansion, modelled with a power law function of solar distance (\sect{Exp_model}), does not provide the quadratic term present in the observations.  Indeed, this weak quadratic term of this expansion model is expected to be masked by the one implied by the interactions with the surrounding medium, \ie\ compression from the sheath and/or from an overtaking stream (like in the examples of \figS{Vfit}a,c,e).
   
We conclude that the velocity fits, especially the quadratic fit, provide a fair representation of the data.  In particular,  typically the residuals remain small compared to the global velocity variations $\dVlin$ and $\dVquad$ (\fig{dV}).  Finally, we notice that the examples of \fig{Vfit} show typical values of $\sdRes$, so they provide fair examples of the typical fluctuations present in the studied MCs.
      
\begin{figure}[t!]        
\centering
\includegraphics[width=0.5\textwidth, clip=]{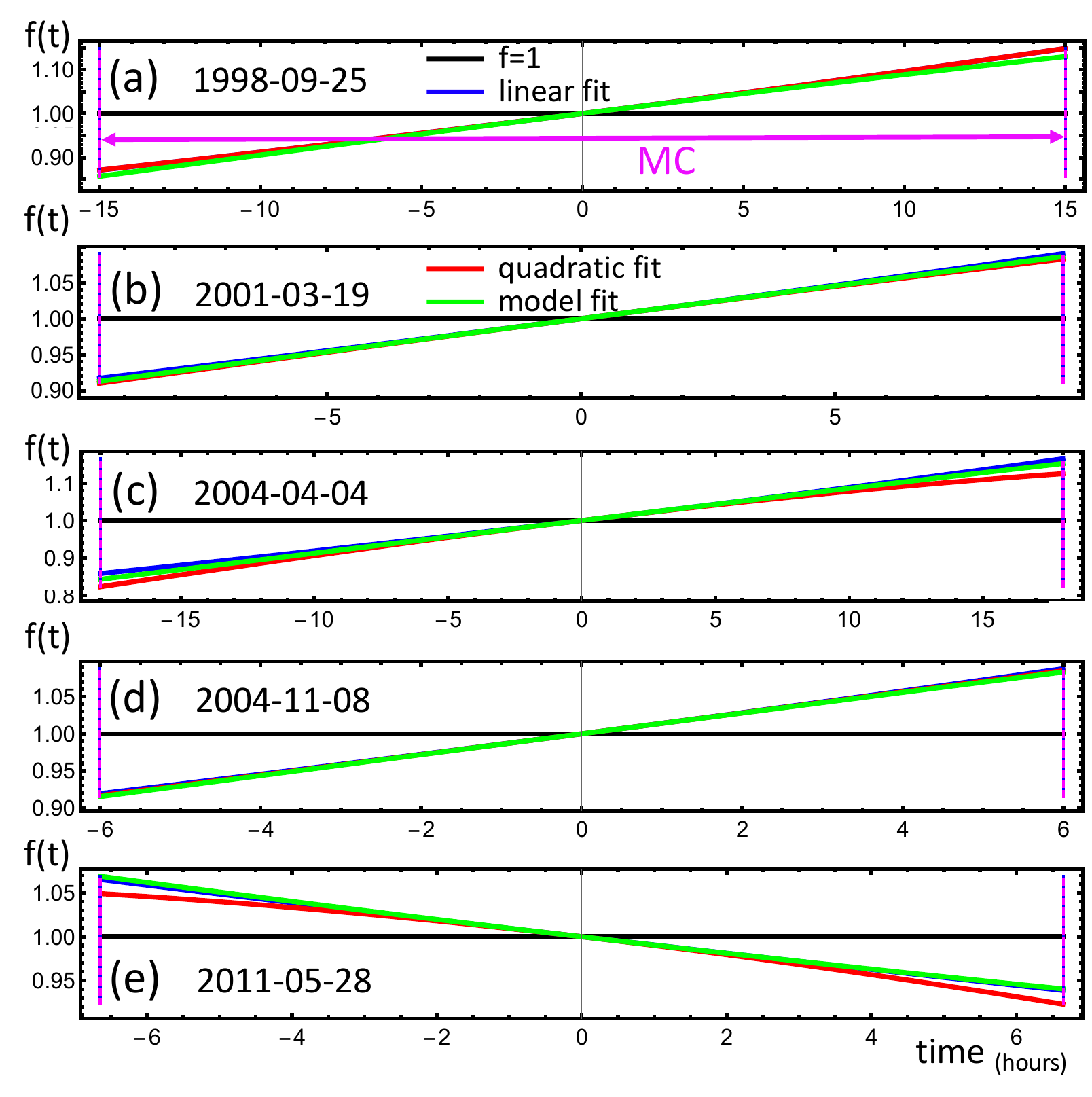}
\caption{Expanding factors $\ffit (t)$, \eq{fFit}, corresponding to the MC velocities shown in \fig{Vfit}.
$\ffit (t)$ are derived from linear, quadratic and model fits of the observed velocities.  
The temporal MC centre, located at t=0, is the reference time for correcting the aging effects (\ie\ $f(0)=1$).
}
 \label{fig_f(t)}
\end{figure}

\subsection{Expansion factors}
\label{sect_f_profiles}
The expansion factor is derived from \eq{fFit} with the coefficients $a,b,c$ provided by the velocity fits.  The central time of the MC, $\tc $, is selected for the reference time then $\ffit(\tc)=1$. 
The five selected MCs have a nearly linear variation of $f(t)$ with time with at most small deviations between the linear and quadratic fits as shown in \fig{f(t)}.  
The small contribution of the quadratic term in $f(t)$ is partly a consequence of its low contribution in the velocity profile compared to the linear term (\fig{dV}a). 
Next, the quadratic term of $f(t)$, in \eq{fFit}, is divided by a factor 2 compared to $\Vfit (t)$ in \eq{Vfit} ($c$ is divided by 2 in \eq{fFit} as a result of the integration within \eq{f(t)int}). This further decreases its contribution in $\ffit(t)$ compared to $\Vfit(t)$.
Finally, the variation of velocity across an MC is typically small compared to its central (or mean) velocity, so that \eq{fFit} can be expanded to the first order in $t-\tc$ to a good approximation, even better than $\Vfit(t)$ (lower quadratic term), providing a nearly linear profile for $f(t)$, as shown in \fig{f(t)}.  

We Furthermore notice that, if the velocity data were directly implemented in \eq{f(t)int} to compute $f(t)$, the time integration would smooth the velocity fluctuations efficiently away from the MC center. Then, we conclude that the expansion factor away from the MC center would be close to the above estimations with data fits and dominated by the linear term when the velocity profile is dominantly linear, within an MC, which is typical.

\begin{figure}[t!]        
\centering
\includegraphics[width=0.5\textwidth, clip=]{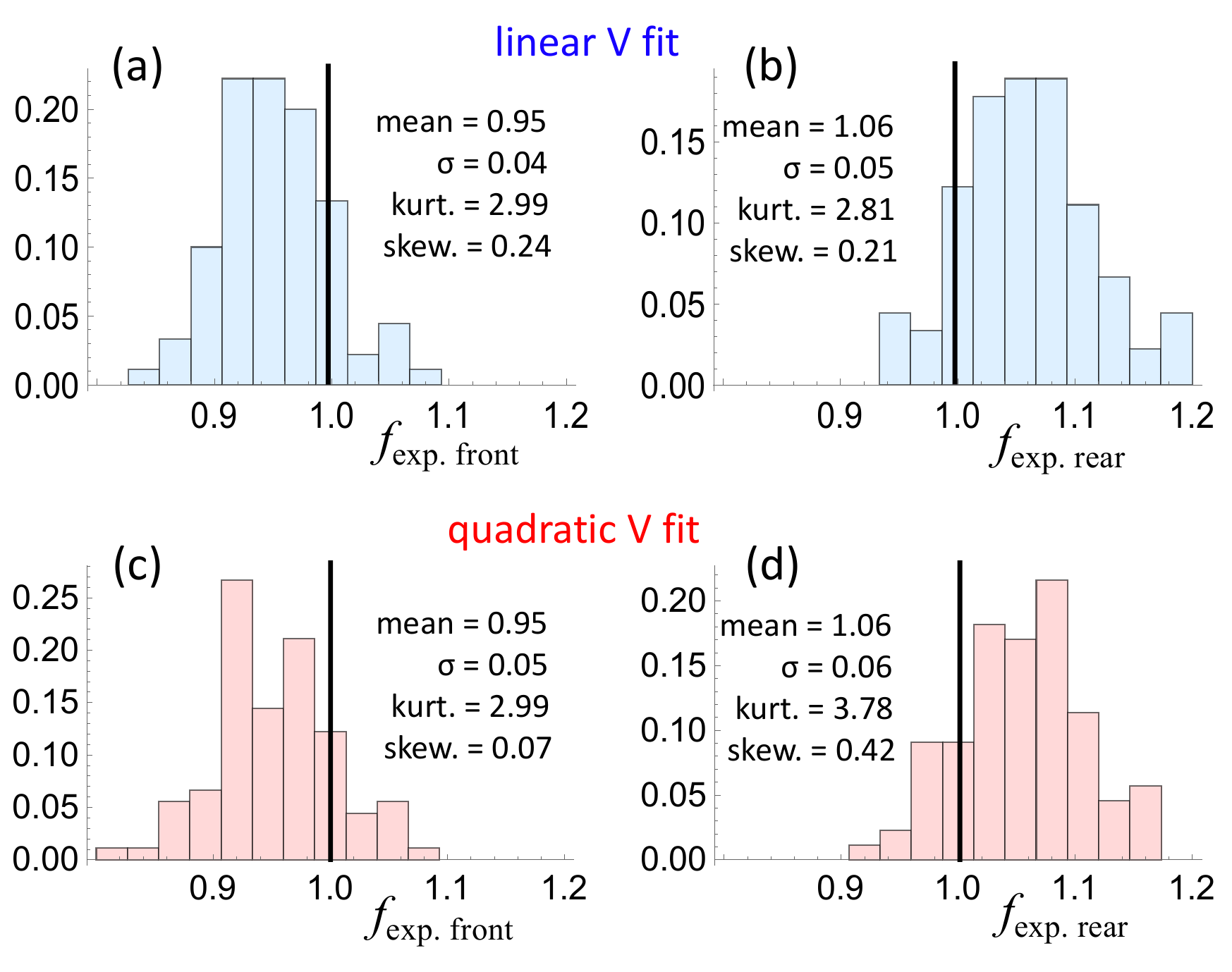}
\caption{Histograms of the expansion factors $\ffit$, \eq{fFit}, computed 
  (a,c) at the front, and
  (b,d) at the rear MC boundaries.
They are the extreme values of $f(t)$ across MCs.  
$\ffit $ values are derived from linear (top) and quadratic (bottom) fits of the observed velocities (similar results are obtained with the model fit).
}
 \label{fig_f_histo}
\end{figure}

The importance of the MC expansion during the spacecraft crossing is shown in \fig{f_histo} with the expansion factors found at the front and rear boundaries.
They are extreme values within each MC (see the analysis after \eq{f(t,Vobs)} and \fig{f(t)}).
All other parameters being equal, more extreme values of $\ffit$ values, \ie\ away from unity, are expected at the periphery of the longer duration MCs like obtained for the MCs in \figS{f(t)}a,c  \citep[see][for a further theoretical analysis]{Demoulin08}.  
      
   The expansion factors are spread in the interval $[0.8,1.1]$ for the front values, and $[0.9,1.25]$  for the rear values.  This shows that the expansion weakly transforms the MCs during their observing time, as expected from the typical values (0.9, 1.1) derived at the end of \sect{Exp_model}.
   The histograms with linear and quadratic fits are slightly different indicating that a fraction of MCs have some differences in $f_{\rm exp. front}$ and $f_{\rm exp. rear}$, like in the examples of \figS{f(t)}c,e.  However, the statistical parameters of $f_{\rm exp.}$ distributions are close, and globally there are only weak differences in the expansion factors computed with linear and quadratic fits. 

A few MCs have $f_{\rm exp. front}>1$ or $f_{\rm exp. rear}<1$ indicating a compression (\eg\ 11\% have $f_{\rm exp. front}>1$ and 18\% have  $f_{\rm exp. rear}<1$ with a quadratic velocity fit).
Still for most MCs, the expansion factor is typically below (resp. above) unity at the front (resp. rear) boundary, respectively, showing that MCs are typically in expansion.

\begin{figure}[t!]        
\centering
\includegraphics[width=0.5\textwidth, clip=]{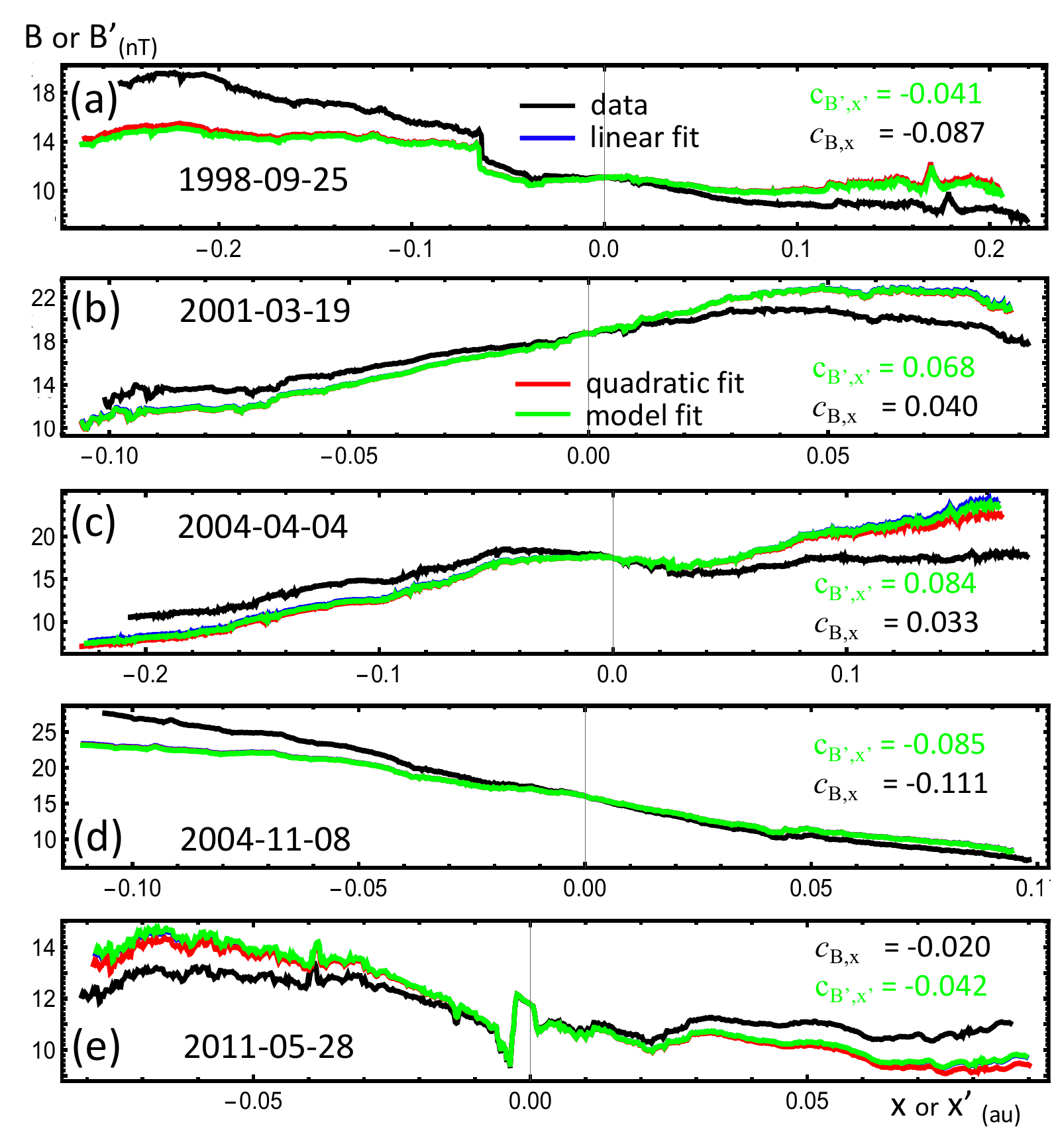}
\caption{Magnetic field magnitude, $B$, across the same MCs as shown in \figs{Vfit}{f(t)}.
The black lines show the original data versus the spatial coordinate $x$, \eq{x(t)}, along the spacecraft trajectory. 
   The coloured curves show the results of removing the aging effect both on $B$ and $x$ with the expansion factors shown in \fig{f(t)}, so they show $B'(x')$ (\eqs{x'}{B'}). The plotting order is black, blue, red and green lines, and the model fit curves mostly mask the underlying results obtained with other fits. 
   A couple of magnetic asymmetry values, $\cB{B}{x}$ and $\cB{B'}{x'}$, are reported on the right side of each panel (with the same colour convention).
}
 \label{fig_Bcor}
\end{figure}

\section{Magnetic asymmetry and expansion rate of MCs} 
\label{sect_Exp_Asy}

\subsection{Magnetic asymmetry}
\label{sect_B_Asymmetry}

The asymmetry of the magnetic field intensity in MCs was quantified by \citet{Janvier19} and \citet{Lanabere20}.   They use the coefficient $\cB{B}{t}$ defined as
   \BE \label{eq_cBt_def}
    \cB{B}{t} = \int_{\tstart}^{\tend} \frac{t-\tc}{\tend-\tstart} \, B(t)\, \rmd t
       \;\bigg/ \int_{\tstart}^{\tend} B(t)\, \rmd t \,,
   \EE
with the central time $\tc = (\tstart + \tend)/2$.

We quantify the $B(x)$ profile asymmetry in a similar way by defining $\cB{B}{x}$ like $\cB{B}{t}$, but with the integration done on the spatial coordinate $x$. 
    \BE \label{eq_cBx_def}
    \cB{B}{x} = \int_{\xstart}^{\xend} \frac{x-\xc}{\xend-\xstart} \, B(x)\, \rmd x
       \;\bigg/ \int_{\xstart}^{\xend} B(x)\, \rmd x \,,
   \EE
with the central position $\xc = (\xstart + \xend)/2$.  The normalisation by the MC size $(\xend-\xstart)$ at the denominator implies that $\cB{B}{x}$, like $\cB{B}{t}$, is dimensionless.
$x$ is computed with \eq{x(t)} and could be further corrected from the aging effect (with $x$ replaced by $x'$, see \sect{Rem_Method}).  

When the $B(x)$ profile is symmetric around $\xc$, $\cB{B}{x} =0$.  $|\cB{B}{x} |$ increases as the asymmetry of the profile increases, with $\cB{B}{x}$ negative when $B$ is stronger before $\xc$, and positive when the field is more concentrated toward the MC rear. $\cB{B}{x}$ is the difference of two oppositely signed quantities (for $x<\xc$ and $x>\xc$), more over it includes the normalisations by the spatial size and the full integral of $B$.   All these contribute to define small values of $|\cB{B}{x}|$ while the asymmetry of $B(x)$ in a studied MC may appear very significant like shown in plots of $B(x)$, such as in \fig{Bcor} (black curves).

   The expected range of $\cB{B}{x}$ values could be computed with the simple model where $B(x)$ decreases linearly by $\Delta B \geq 0$ across the MC with a central field $\Bc$. Including this model in \eq{cBx_def}, we derive $\cB{B}{x} = -\Delta B /(12 \Bc)$.  
   An extreme case is when $B=0$ at the rear boundary, which corresponds to $B=2\,\Bc$ at the front boundary, then $\Delta B = 2\,\Bc$.  For this strong $B$ asymmetry, $\cB{B}{x} =- 0.167$.   The MC data shown in \fig{Bcor}d (black line) are comparable to this profile, while slightly less asymmetric. 
   If we rather include $\Delta B = \Bc$ in the model, so a magnetic field decreasing from $3 \Bc /2$ at the front to $\Bc /2$ at the rear, $\cB{B}{x} \approx - 0.083$.  This result is close to the MC data shown in \fig{Bcor}a (black line). 
   
The coefficients $\cB{B}{x}$ and $\cB{B}{t}$ could be compared by performing a Taylor expansion of $\Vobs(t)$ and $B(t)$.  This allows us to compute analytically the integrals in \eqs{cBt_def}{cBx_def}.
For our purpose, linear expansions of $\Vobs(t)$ and $B(t)$ are sufficient within MCs. This implies  
     \begin{eqnarray}
     \cB{B}{t} &=& db\,/12 \,, \label{eq_cBt} \\
     \cB{B}{x} &=& \cB{B}{t}\, (1-dv^2 /20) \,/\,(1+db \, dv /12)  \,, \label{eq_cBx(cBt)}
     \end{eqnarray}
where $db= \Delta B /\Bc$ and $dv= \Delta \Vobs /\Vc$ are the relative changes across the full MC.
In the analysed MCs we have $|db|\lesssim 1$ and $|dv|\lesssim 0.3$ (\eg\ see $\dVlin$ in \fig{dV}). 
Within these limits, \eq{cBx(cBt)} implies $\cB{B}{x} \approx \cB{B}{t}$ with the largest difference coming from the denominator contribution.  The computation could be extended to more terms, still the low values of $|dv|$ imply even smaller contributions. 

In summary, $|\cB{B}{x}|$ increases with the magnetic asymmetry, and a large asymmetry is marked with $|\cB{B}{x}|$ around or larger than $0.1$. 
$\cB{B}{x}$ is close to $\cB{B}{t}$ for MCs and $\cB{B}{x}$ includes the effect of aging and the intrinsic spatial asymmetries.   $\cB{B}{x}<0$ marks a $B$ field stronger in the MC front, and $\cB{B}{x}>0$ marks a $B$ stronger at the rear.

\begin{figure}[t!]        
\centering
\includegraphics[width=0.5\textwidth, clip=]{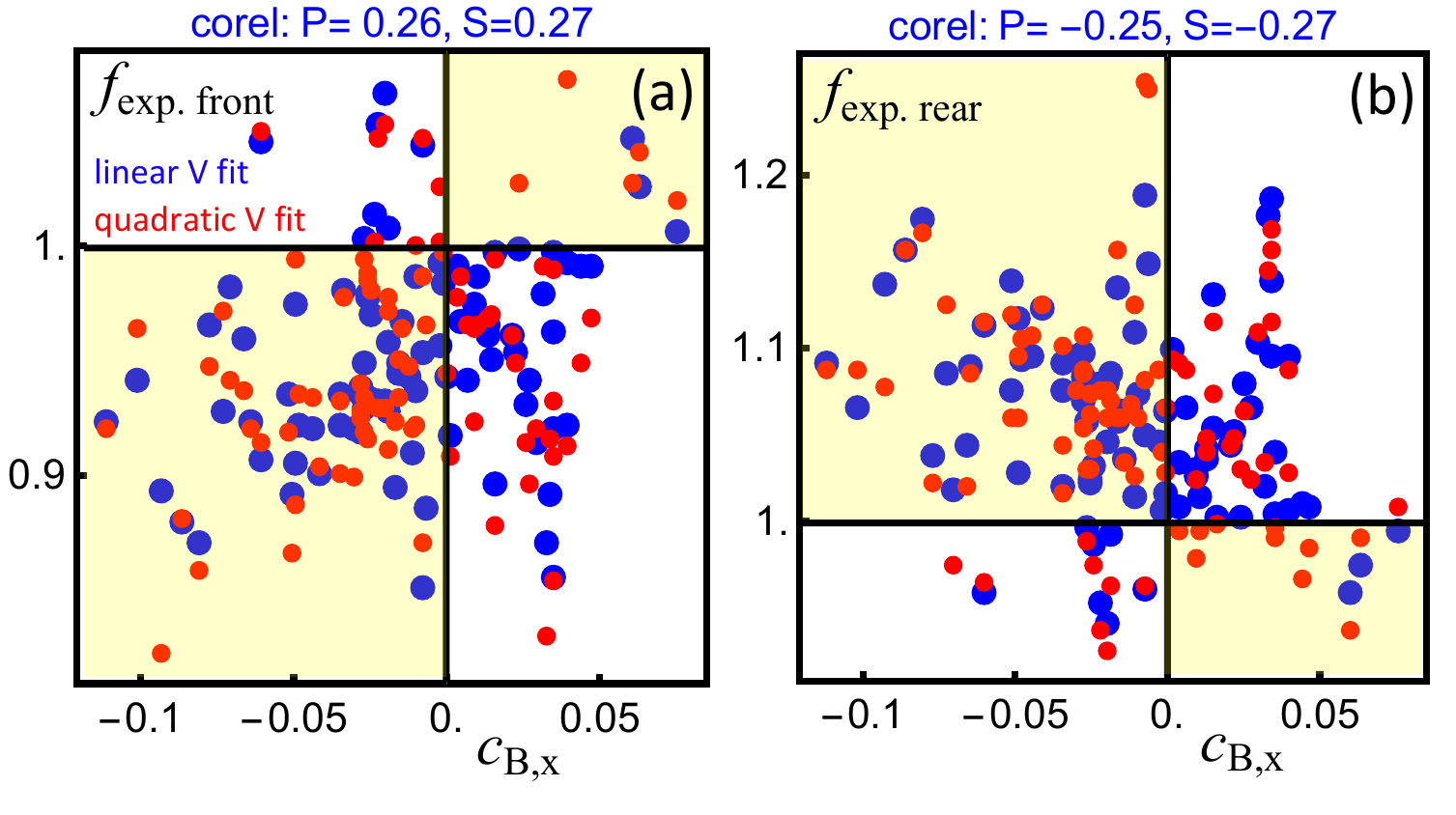}
\caption{
Expansion factors (a) at the front and (b) at the rear MC boundaries as a function of the magnetic asymmetry coefficient $\cB{B}{x}$, \ie\ \eq{cBx_def} applied to the magnetic data.  The results of linear (blue) and quadratic (red) fits for a given MC are shown with nearby couple of points located on the same vertical line.  The model fit derived from \eq{VmodQuadra} (in green in earlier figures), is omitted because of nearly redundant results. 
The Pearson (P) and Spearman (S) correlation coefficients are reported on the top of each panel for linear velocity fits.
 The aging effect alone is expected to create magnetic asymmetry correlated with the expansion factors, so that for asymmetry due to aging, data points are expected to be included in the yellow regions.}  
 \label{fig_fexp_CB}
\end{figure}

\subsection{Effect of aging effect on magnetic asymmetry}
\label{sect_Effect_Expansion}

We report in \fig{fexp_CB} the values of $f(t)$ found at MC boundaries with the asymmetry of $B$ computed from \eq{cBx_def}. 
This confirms the results of histograms (\fig{f_histo}) that linear and quadratic fits show moderate differences in expansion factors as shown with the typical small shift in ordinate between blue and red pair of points (only a few cases have a difference of $f_{\rm exp. front}$ and $f_{\rm exp. rear}$ between 0.05 and 0.1).
 
Next, we explore the expected effect of aging on the observed $B$ field. 
We recall that $f_{\rm exp. front}<1$ and $f_{\rm exp. rear}>1$ is the expected signature of the expansion, and the opposite inequalities are for contraction.
Next, let us consider an hypothetical intrinsic symmetric MC ($\cB{B}{x} = 0$) like considered in \citet{Demoulin08} with a FR model.  The inclusion of expansion implies $\cB{B}{x}<0$, while compression implies $\cB{B}{x}>0$ on the data simulating a spacecraft crossing the FR in the conditions of observations (so including the aging effect).  

With a significant effect of the aging effect on magnetic asymmetry, we would expect a clustering of the data points in the yellow regions in \fig{fexp_CB}.
Even more, larger $|\cB{B}{x}|$ values are expected when expansion or compression is more important.  Such expectations are not present in \fig{fexp_CB} since the points are dispersed, both globally and even inside the yellow regions. 
We conclude that the results of \fig{fexp_CB} imply that the aging effect is not the main cause of the magnetic asymmetry observed in MCs.  Still, since moderate correlation coefficients are found ($\approx \pm 0.26$, top of \fig{fexp_CB}), the aging effect is expected to have a moderate effect on the asymmetry.

\section{Removing the aging effects within MCs} 
\label{sect_Rem}

The main aim of this section is to remove the bias produced by coupling space and time when single point observations are made on different elements of plasma at different time (aging effect). Then, the aim is to provide data like if they were obtained at the same time within the 1D-cut of the observed MC. 

\subsection{Method to correct the aging effects}
\label{sect_Rem_Method}

The estimation of $f(t)$ presented above allows us to correct the observations of the aging effect,  within the hypothesis of a self-similar expansion.  More precisely, from the informations derived from the observed velocity profile, we correct the observed magnetic profile $\vec{B}(t)$ in field strength, then we transform it to $\vec{B}(x')$ like if the observations were done at the same time across the MC. 
The corrections are both on the spatial scale and on the field strength as follows. 

The correction on the spatial scale, by $1/f(t)$, is to be applied on the elementary length $\rmd x(t)$, \eq{dx(t)}, observed at time $t$ (and not on $x(t)$, \eq{x(t)}, which cumulates elementary length observed at different times). This defines the spatial coordinate $x'$ like if the full MC would have been observed at the time $\tc$ with 
  \BE \label{eq_x'}
    x'(t) = \Int{\tref}{t} \frac{\Vobs(\ti)}{f(\ti)} \,\rmd \ti  \,.
  \EE
Since $f(t)$ is mostly a linear function (\fig{f(t)}), this implies an antisymmetric increase (resp. decrease) of the front (resp. rear) extension of the MCs in expansion as shown in \fig{Bcor}a-d (same MC examples as in \figs{Vfit}{f(t)}). 
A reverse effect is present for MCs in compression (\eg\ \fig{Bcor}e).  This transformation of $x$ to $x'$ almost conserves the MC size because the front is extended by nearly the same amount than the rear is contracted (to second order in the MC duration $\delta t$).

The correction of the magnetic field needs another hypothesis since the data only allow us to derive the expansion along the spacecraft trajectory.  The 3D expansion could be derived from observations of the same MC by two spacecraft located at different radial distance from the Sun, like performed for one MC by  \citet{Nakwacki11}.  The results are close to isotropic self-similar expansion.  A nearly isotropic expansion is also generically expected with an expansion driven by the adjustment towards total pressure equilibrium between the MC and its surroundings when the MC moves away from the Sun \citep{Demoulin09}.
Furthermore, the analysis of \insitu\ observations indicates an expansion rate, $\zeta$, in the radial direction (away from the Sun) comparable on average to the expected ortho-radial expansion rate, $\approx 1$. This implies spatial scales in three orthogonal directions nearly proportional to solar distance \citep{Demoulin08,Gulisano10,Gulisano12}.      
      
Following the results above on expansion, we suppose an isotropic self-similar expansion to correct the magnetic field.    
This evolution can be included in the more general context of ideal MHD for size rescaling of an initial MHD state $\vec{B}_0(\vec{r})$ as $\vec{B}(f \vec{r}) = f^{-2} \vec{B}_0(\vec{r})$. This re-scaling is the consequence of flux conservation at the level of each elementary fluid bubble.  Indeed, an isotropic expansion (resp. compression) of the fluid between two states means that all spatial scales increase (resp. decrease) like $\vec{r} \rightarrow f \vec{r}$.  This implies that the magnetic field is modified to $\vec{B}_0(\vec{r}) \rightarrow f^{-2} \vec{B}_0(\vec{r})$ to conserve the magnetic flux ($B_0\,\rmd r^2 = f^{-2}\, B_0 \, f^2\, \rmd r^2$).
Here, the rescaling $f(t)$ is function of time (since the MC evolves when the spacecraft crosses it).  Then, in order to remove the aging effect, all magnetic field components are rescaled by multiplying them with $f^{2}(t)$. 
Including the transformation of $t$ to $x'$, \eq{x'}, this provides  
  \BE \label{eq_B'}
  \vec{B}'(x') = \vec{B}(t) \, f^2(t) \,.
  \EE
This transformation ensures magnetic flux conservation. 

The isotropic expansion could be a coarse hypothesis for MCs which are deformed during their propagation in the solar wind, such as shown in some numerical simulations \citep[\eg\ ][]{Cargill02,Manchester04,Lugaz05b,Xiong06}.
However, the shape of FRs is a consequence of the full evolution from the Sun to the spacecraft while the expansion correction is here only applied during the MC crossing, so our hypothesis of isotropic self-similar expansion is expected to be a good approximation during the spacecraft crossing.  Next, while equations, with a different expansion rate in three orthogonal directions have been developed \citep{Demoulin08}, the typical observations by one spacecraft do not allow us to constrain the three expansion factors.   Finally, in view of the moderate corrections introduced by removing isotropic expansion effects, including 3D expansion effects is expected to be a correction at next order of magnitude.

\subsection{Magnetic asymmetry corrected for the aging effect}
\label{sect_Rem_Asymmetry}

The spatial magnetic profiles derived directly from the data and those corrected for the aging effect are compared in \fig{Bcor}, for the same MC examples shown in \figs{Vfit}{f(t)}.  The MC examples have moderate to large $B$ asymmetries and of both signs.  We first describe the results for the MC examples having the two largest aging effect (\figS{f(t)}a,c).
The maximal correction of aging is about 4 nT for $B \approx$ 19 nT (\fig{Bcor}a) and 6 nT for $B \approx$ 17 nT (\fig{Bcor}c), so at most a correction of 35\%.  
The correction reduces the asymmetry of $B$ between the front and the rear of the MC in \fig{Bcor}a, as summarised by the significant reduction of $|\cB{B}{x}|=0.087$ to $|\cB{B'}{x'}|=0.042$.
  The correction is smaller in \fig{Bcor}d ($|\cB{B}{x}|$ is reduced by 25\%). This is especially true at the MC rear where the correction of $x$ shifts the nearly linear $B$ profile toward the MC front, so that $B'(x')$ is close to $B(x)$ in this rear region. The same effect is present in panel a, but there the aging removal is stronger, so $B'(x')$ and $B(x)$ are further away.

   At the opposite, an increase of asymmetry is present from $\cB{B}{x}$ to $\cB{B'}{x'}$ with a factor 1.7 and 2.7 in \fig{Bcor}b,c, respectively. This is a direct consequence of MCs in expansion since the already weaker $B$ at the front is even weaker after the correction while the opposite happens at the MC rear.  As expected for MCs in compression with a stronger $B$ in their front, as shown in \fig{Bcor}e, the asymmetry is also increased after the correction ($|\cB{B'}{x'}|$ is larger than $|\cB{B}{x}|$ by a factor 2.1).
   All these are indications that the aging effect is a weak source of the asymmetry of the magnetic field in MCs, and we quantify this below.  
   
\begin{figure}[t!]    
\centering
\includegraphics[width=0.5\textwidth, clip=]{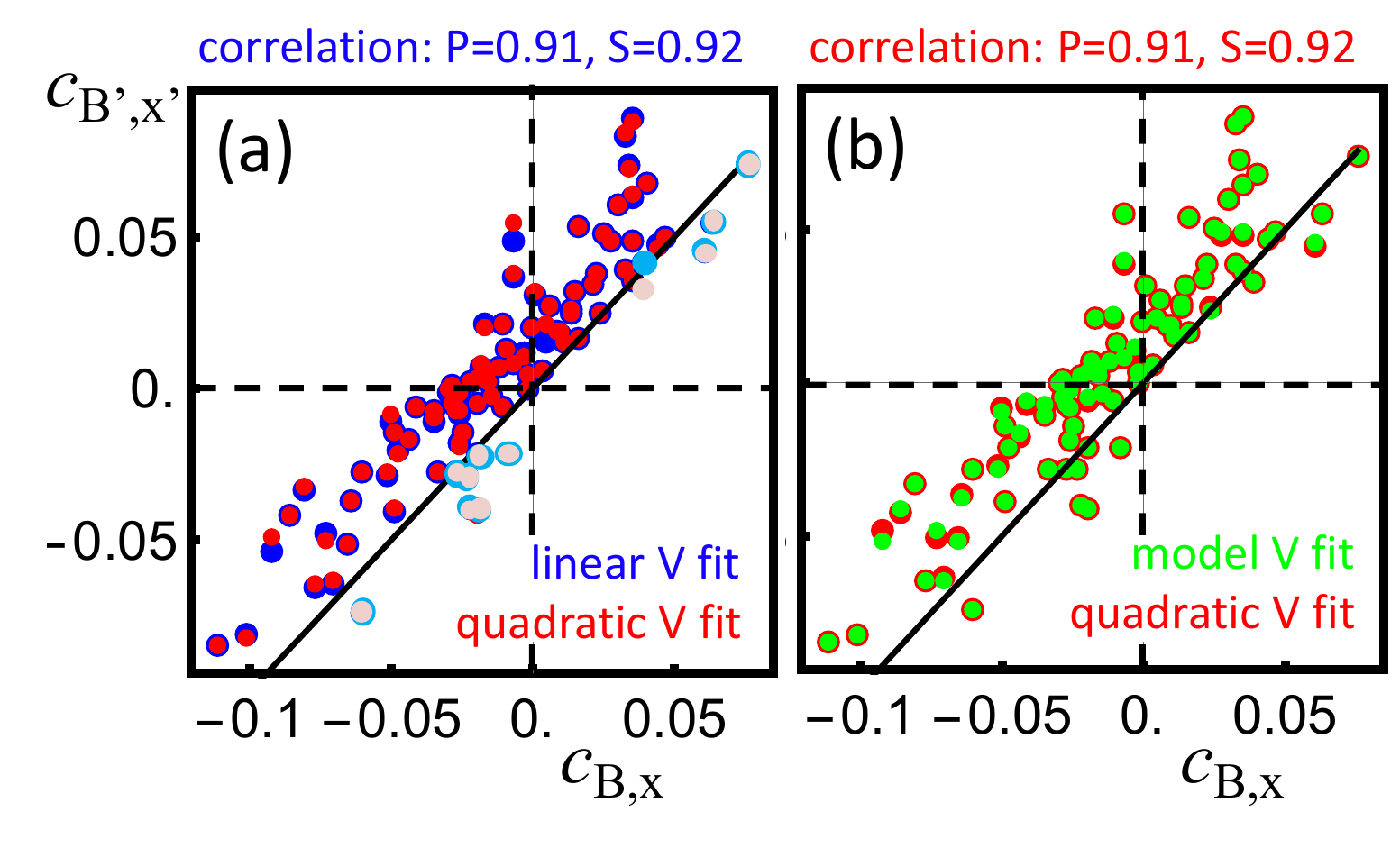}
\caption{
Comparison of the asymmetry coefficient $\cB{B}{x}$ with the aging effect removed (ordinate) as a function of $\cB{B}{x}$ derived from the uncorrected magnetic data (abscissa). 
  (a) The results of linear and quadratic fits of the observed velocity are very similar.  The points with a lighter colour are in compression ($\dVlin >0$, see \fig{dV}a).
  (b) The results of the model fit, \eq{VmodQuadra}, of the observed velocity are almost the same than the results with the quadratic fit.  The diagonal black line is $\cB{B'}{x'}=\cB{B}{x}$.
The Pearson (P) and Spearman (S) correlation coefficients are reported on the panel top for linear and quadratic velocity fits. } 
 \label{fig_CB}
\end{figure}

More generally, \fig{Bcor} provides examples of the following results.
  First, correcting the aging effect reduces the asymmetry for MCs with stronger field in the front, \figS{Bcor}a,d but does not remove it as quantified by the $\cB{B'}{x'}$ values.  A full asymmetry removal of the $B$ asymmetry would require an expansion rate fully incompatible with the observed velocity profile, while the velocity fits are very close to the data (\fig{Vfit}). 
   Second, for some MCs, like in \figS{Bcor}b,c, the removal of the aging effect rather increases the $B$ asymmetry as quantified by the increase of $\cB{B'}{x'}$ compared to $\cB{B}{x}$.
    Finally, the aging corrections have comparable results with the three type of expansion estimations (colour curves are nearly superposed in \fig{Bcor}), as expected with the results of \fig{f(t)}.

The behaviours shown in the examples of \fig{Bcor} are present in most MCs as shown in \fig{CB} where the values of $\cB{B'}{x'}$, with aging removed, are plotted as a function of the $\cB{B}{x}$ values.   
   Removing the aging effect implies that $\cB{B'}{x'}$ is typically shifted by a positive value since the large majority of points are above the diagonal ($\cB{B'}{x'}=\cB{B}{x}$, black line), as expected since most MCs are in expansion (\fig{f_histo}).  For MCs in expansion and with a stronger field at the rear ($\cB{B}{x}>0$) this implies an increase of the magnetic asymmetry ($\cB{B'}{x'}>\cB{B}{x}$), while it is the opposite for MCs with a stronger field in the front ($\cB{B}{x}<0$).
   
   For a minority of MCs in contraction (with $\dVlin >0$ and marked with lighter colours in \fig{CB}a), $\cB{B'}{x'}$ is slightly shifted by a small negative value compare to $\cB{B}{x}$ (points are below the black diagonal).  We also notice that no MC is present in the lower right quadrant of \fig{CB} panels, which is a consequence of a weak contraction and for only a few MCs. 
Finally, $\cB{B'}{x'}$ values are at variable distances from the diagonal, then the correction of $\cB{B}{x}$ is of variable magnitude and independent of the original value of $\cB{B}{x}$.  

The correction of $\cB{B}{x}$ for the aging effect is weakly dependent of the type of velocity fit, since blue, red and green points nearly overlap in Fig. 7.  These similarities are even stronger than in \fig{fexp_CB}.  Indeed, \fig{fexp_CB} shows the most extreme expansion factors.  The difference of $f(t)$ between the three types of velocity fits is lower in the MC core (\fig{f(t)}). 
  More over, $\ffit (t)$ is the product of the contributions from the linear and quadratic terms as shown in \eq{fFit}.  This implies that the quadratic contribution to $\ffit (t)$ is the same in the MC front and rear for the same time difference with the MC centre. We conclude that the quadratic term of the velocity fit is correcting similarly both MC sides, so it has a weak effect on the asymmetry, and in particular on $\cB{B'}{x'}$. 
This implies that the global effect of removing the aging effect on $B(x)$, as computed by $\cB{B'}{x'}$, is comparable for the three types of velocity fits.

Most important, \fig{CB} shows that, in general, removing the aging effect in MCs does not bring the magnetic field to a more symmetric configuration (\ie\ $\cB{B'}{x'}$ closer to 0).  Rather $\cB{B'}{x'}$ values are very closely correlated with the original $\cB{B}{x}$ values as shown both with Pearson and Spearman correlation coefficients (top of \fig{CB}).  We conclude that the aging effect is in general a small contribution to MC asymmetry except for a cluster of points near $\cB{B'}{x'}=0$.

\begin{figure}[t!]    
\centering
\includegraphics[width=0.5\textwidth, clip=]{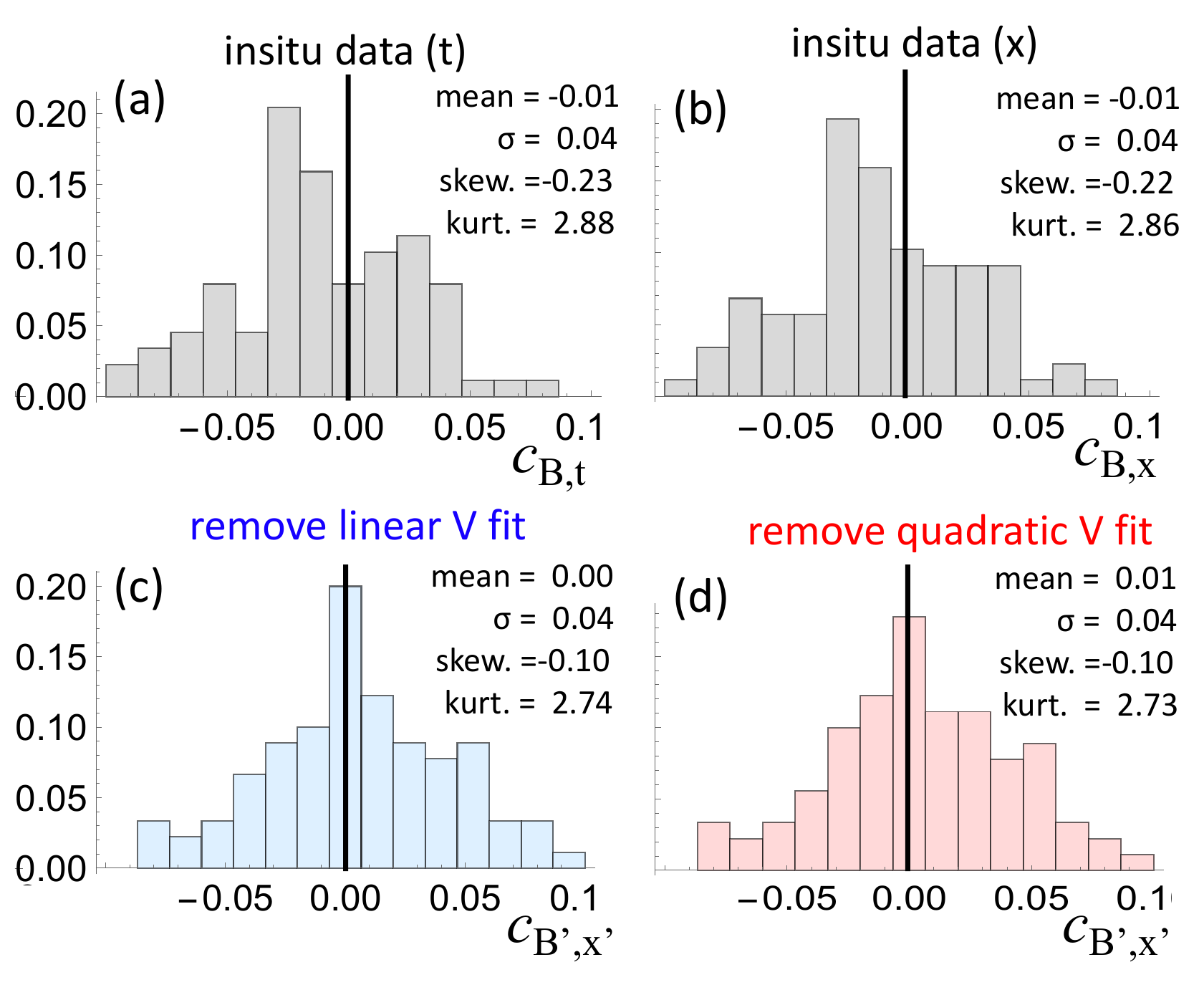}
\caption{Comparison of the histograms of the asymmetry coefficients.
(a,b) $c_{\rm B}$ is computed directly with the magnetic data function of (a) the time $t$ and (b) the spatial coordinate $x$ (\eq{x(t)}, $x$ includes aging effect).  
(c,d) $\cB{B'}{x'}$ is computed after removing the aging effect on the spatial coordinate ($x'$, \eq{x'}) and on the magnetic field (\eq{B'}).  The expansion factor, $f(t)$, is derived with (c) a linear fit, and (d) a quadratic fit of the observed velocity (\eq{Vfit}).}
 \label{fig_CB_histo}
\end{figure}

\subsection{Distribution of magnetic asymmetry}
\label{sect_Rem_Distribution_cB}

Histograms of \fig{CB_histo} confirm the previous results.
  First, the transformation of abscissa from $t$ to $x$ only weakly changes the asymmetry (\figS{CB_histo}a,b) in agreement with the earlier results from \eqs{cBt}{cBx(cBt)}. 
We notice that all $c_{\rm B}$ values reported in \citet{Lanabere20} are larger by a factor 2. This rescaling does not change any of their conclusions.
 Next, the correction of the aging effect globally shifts $\cB{B'}{x'}$ to slightly more positive values than $\cB{B}{x}$, while in general this does not decrease significantly the values of $|\cB{B'}{x'}|$ (similar standard deviation $\sigma$, and similar distribution shape as quantified by the skewness and the kurtosis).  More over the results are robust since the $\cB{B'}{x'}$ distribution is only weakly affected by the method used to remove the aging effect (\figS{CB_histo}c,d), and the histogram of the model V fit is similar to the quadratic fit.
 
Histograms also reveal points which are not outstanding in \fig{CB}, as follows. 
The distribution of $\cB{B}{x}$ is peaked around $\approx -0.025$, \fig{CB_histo}a,b (even if the kurtosis is comparable to the value of 3 for a normal distribution).  Removing the aging effects strengthen and shift this peak to $\cB{B'}{x'} \approx 0.$ (\figS{CB_histo}c,d). This outstanding peak represents globally symmetric $B(x)$ profiles.  Indeed, 28\% of the MCs have $|\cB{B'}{x'}|<0.01$.
These MCs are expected to be closer to theoretical FR models, which are typically symmetric.
Then, for this subset of MCs, the magnetic asymmetry in observations is mostly the result of aging effect.

The distribution of $\cB{B'}{x'}$, corrected for the aging effect, is almost symmetric with both low mean and skewness.  This is a surprising result since the physical mechanisms creating $\cB{B'}{x'}<0$ or $\cB{B'}{x'}>0$ are expected to be different.  For example, a stronger total pressure in the MC sheath is expected to imply $\cB{B'}{x'}<0$ while a fast overtaking stream at the MC rear is expected to imply $\cB{B'}{x'}>0$. Next, the bending of the FR axis, concave towards the Sun, is expected to increase $B$ at the rear of the MC compared to its front, then to increase  $\cB{B'}{x'}$.   All these mechanisms, as well as their magnitudes, are expected to be independent, so they are expected to contribute differently to the magnetic field asymmetry.  Then, an asymmetric distribution of $\cB{B'}{x'}$ is rather expected,
in contrast with the results in \fig{CB_histo}. 

\section{Conclusion} 
\label{sect_Conclusion}

With coronagraphic and heliospheric imagers, ICMEs are generally observed to expand when they move away from the Sun, while \insitu\ observations confirm this with direct measurements of the proton bulk velocity.  However, in contrast to imagers, \insitu\ observations are done at various times during the spacecraft crossing, coupling spatial shape with time evolution, so that the measurements are directly affected by the aging effect.  The main aim of this study is to estimate this so-called aging effect, then to do a corresponding correction of measurements in order to provide data like if they were observed at the same time along the full crossed structure.  We apply the developed method to magnetic clouds (MCs) of quality 1 and 2 in Lepping's list in order to study only the best observed cases (crossing closer to the flux rope (FR) core, stronger magnetic field, and less perturbed cases).  This selection is expected to provide the clearest results. 

The measured \insitu\ velocity along the spacecraft trajectory is decomposed in a global and an expansion contributions.   We justify that the global velocity is nearly constant during the crossing of an ICME at 1 au (\ie\ its limited change cannot explain the observed \insitu\ variations of the velocity).  Then, with the hypothesis of self-similar expansion during the spacecraft crossing, we derive a generic relation which expresses the expansion factor as a function of the observed velocity.   With the observed duration of MCs, we show that a Taylor expansion of the velocity up to the second order is sufficient for applications to MC data.   Then, the observed velocity is fitted with either a linear, either a quadratic, function of time to filter the local fluctuations. Finally, the corresponding expansion factors, as a function of time, are derived along the spacecraft trajectory.  We also derive the expansion factor with a model which supposes a power law evolution of the MC size with solar distance and the free parameters are determined by a fit to the velocity data.
   
The spatial coordinate $x$ along the spacecraft trajectory is computed by a temporal integration of the observed velocity. This converts time to space for each parcel of plasma, so it adds the sizes of plasma blobs observed at different times.  Then, $x$ is not a true spatial coordinate at a given time since it includes the expansion of the configuration.  The derived expansion factor allows us to correct $x$ for the aging effect, to derive the coordinate $x'$, \eq{x'}, like if the whole MC was observed at the same time, that we set at the observed central time.   Next, we correct the magnetic field components with \eq{B'}, which provides $\vec{B'}(x')$.  Both corrections assume a self similar expansion of the MC.  Then, this study allows us to both quantify the importance of the aging effect, and to remove it from the observed $\vec{B}(t)$ profile to finally deduce the spatial $\vec{B'}(x')$ variations like if the observations were done at the same time within the MC, \ie\ without aging effect.
   
The shapes of $B(t)$, $B(x)$, and $B'(x')$ profiles are quantified with the asymmetry parameters $\cB{B}{t}$, $\cB{B}{x}$, and $\cB{B'}{x'}$, respectively (defined by \eqs{cBt_def}{cBx_def}).
The values $\cB{B}{t}$ and $\cB{B}{x}$ in MCs are close by and their histograms show a shift to negative values which reflect, on average, stronger $B$ values in the front region of MCs.  The histogram of $\cB{B}{x}$ is slightly transformed to the one of $\cB{B'}{x'}$ by removing the aging effect, with very similar results for the three types of velocity profiles fitted to the data.  The main change is the presence of a strong peak around $\cB{B'}{x'}=0$, so globally symmetric $B'(x')$ profiles (\fig{CB_histo}).  For this subset of MCs, about one fourth of the studied set, the aging effect is the main source of the observed $B(t)$ asymmetry.
  Next, for the fraction of MCs (about 22 \%) 
both in expansion and having a stronger fields at the rear ($\cB{B}{x}>0$), removing the aging effect rather increases the asymmetry.
   For the remaining MCs ($\cB{B'}{x'} \lesssim 0.03$), removing the aging effect leads to magnetic profiles slightly more symmetric. Still, a global symmetric $B$ ($\cB{B'}{x'}\approx 0$) would require an expansion rate much stronger and so incompatible with the observed velocity profile.   

In summary, removing the aging effect does not bring $|\cB{B'}{x'}|$ in general closer to zero than $|\cB{B}{x}|$ since both the dispersion and the wings of $\cB{B}{x}$ and $\cB{B'}{x'}$ histograms are similar. 
We conclude that the aging effect is not the main origin the observed $B(t)$ asymmetry for \insitu\ data of MCs. Several sources of intrinsic magnetic asymmetry are possible, in particular a higher compression by the surrounding medium on one MC side, either by a strong sheath at the front or an overtaking fast stream at the rear.
  
Finally, while the aging effect is typically weak, it is still worth to correct its effects in particular on large events (where the effects are larger).  This decouples the time evolution from the spatial magnetic configuration of MCs.
We compare three types of fits of the observed velocity (one linear, one quadratic and one derived from power law model for the size evolution with solar distance).  They imply nearly identical corrections, then the aging effect could be well removed from any of the methods used here.   
These methods provide a spatial profile of $\vec{B}(x)$ similar to the one which would be obtained if the full MC would be observed at the same time as its centre.

Removing the aging effect on the data is a promising alternative to technics fitting both the velocity and magnetic data with an expanding FR model.  First, there is no longer the need to introduce an \textit{ad hoc} coefficient to include both the magnetic field and velocity data in the minimised function.  Second, the number of free parameters is decreased, then more elaborated magnetic models with more free parameters could be used.  Third, the corrected magnetic data can be directly fitted by any static model or analysed by an alternative method (\eg\ with minimum variance analysis or by solving the Grad-Shafranov equation).  This allows us to compare more directly the results derived from several methods.  Finally, the developed method to remove the aging effect can be applied more generally to magnetic ejecta, within ICMEs, assuming that they have an isotropic self-similar evolution during the spacecraft crossing.

\begin{acknowledgements}
 We recognise the collaborative and open nature of knowledge creation and dissemination, under the control of the academic community as expressed by Camille Noûs at http://www.cogitamus.fr/indexen.html and
we thank Bojan Vršnak for his comments which improved the manuscript. 
S.D. acknowledges partial support from the Argentinian grants UBACyT (UBA), and PIP-CONICET-11220130100439CO.  S.D. thanks the LIA project (LIA1208).
This work was partially supported by a one-month invitation of P.D. to the Instituto de Astronom\'ia y F\'isica del Espacio,  
and by a one-month invitation of S.D. to the Observatoire de Paris.
This work was supported by the Programme National PNST of CNRS/INSU co-funded by CNES and CEA. 
S.D. is member of the Carrera del Investigador Cien\-t\'\i fi\-co, CONICET.

\end{acknowledgements}

%
\bibliographystyle{aa}
\bibliography{mc}
\IfFileExists{\jobname.bbl}{}
{\typeout{}
\typeout{****************************************************}
\typeout{****************************************************}
\typeout{** Please run "bibtex \jobname" to optain}
\typeout{** the bibliography and then re-run LaTeX}
\typeout{** twice to fix the references!}
\typeout{****************************************************}
\typeout{****************************************************}
\typeout{}
}


\end{document}